\documentclass[prb,twocolumn,twoside,showpacs]{revtex4}

\newcommand{\factorgr}{1.0}
\newcommand{\factorkl}{0.9}

\usepackage{graphicx}
\usepackage{psfrag}

\begin{document}

% Information for title 

\title[Ab initio investigation of Fe$|$ZnSe(001)]
{Ab initio investigation of structural and magnetic properties of Fe$|$ZnSe(001)}
\date{\today}
\author{H.~Dobler}
\author{D.~Strauch}
\affiliation{Institut f{\"u}r Theoretische Physik, Universit{\"a}t
  Regensburg, 93040 Regensburg, Germany}

% Abstract

\begin{abstract}
  We investigate 1, 2, 3, and 4 monolayers of iron grown on the (001)
  surface of ZnSe using spin density functional theory, plane-wave expansion,
  and ultrasoft pseudopotentials. 
  An ideal structure of the interface is plausible due to the very
  small lattice mismatch between both materials.
  However, we observe dramatic deviations from this idealized
  configuration. 
  The most distinctive variations arise in the case of a
  single monolayer. Here a terminating Zn layer is buoying
  upwards. The thicker the Fe layer the weaker are the relaxations in
  the iron region. Nevertheless, in the case of the Se-terminated
  surface, the structure changes significantly inside the semiconductor
  near the interface. The profiles of the electronic density and the 
  magnetization perpendicular to the surface are quite sensitive to the change 
  from the ideal to the real structure.
\end{abstract}


\maketitle

% Introduction

\section{Introduction}

Magnetoelectronics\cite{Pri98} and spintronics\cite{Wol01} are
rapidly growing fields in applied solid state physics. Since
the proposal of a spin field-effect transistor by Datta and
Das\cite{Dat90} ferromagnet--semiconductor (FM--SC) heterostructures have become
very interesting candidates for the realization of spin dependent injection of
electrons in a semiconductive material. Therefore, theoretical\cite{But97,Mac98,Kos00} and experimental\cite{Kre87,Rei00,Ben01} effort has been made. Towards this goal, calculations were made for FM--SC--FM heterostructures, while thin Fe films epitaxially grown on semiconductor surfaces were investigated experimentally. 

The first Fe$|$GaAs(001) systems\cite{Kre87} had shown magnetic dead layers. This phenomenon of Fe layers at the interface between the two materials with nearly vanishing magnetic moment can be prevented by lowering the temperature of the substrate to room temperature during the film growth.\cite{Zol97} Higher temperature leads to a mixing of the materials at the interface which sensitively influences the magnetic properties. Thus, structure plays an important role. 

Nevertheless, the theoretical investigations of FM--SC heterostructures usually assume an ideal interface structure. This is suggested by the very small mismatch of the lattice constants between diamond-structure elemental or zincblende-structure compound semiconductors and of $\alpha$-Fe. Therefore, four Fe cubes excellently fit on a single fcc cube of the semiconductor material. The advantage of these computations is the possibility of considering the contribution of orbital magnetic moments. This is done within the LKKR-method\cite{Mac90} which provides the opportunity to take into account the relativistic spin-orbit coupling. However, the possibility of sensitive reactions of the total magnetization to minor structural changes is fully neglected.

In this article we show that these idealizing assumptions are not realistic in many cases. Therefore, we determine the equilibrium structure of the observed systems and compare the ideal and real density profiles of the valence electrons and of the magnetization. We will see that the deviations from the ideal behavior transcend small quantitative corrections even for the structural results. Because the orbital contribution can be shown\cite{Kos00} to contribute less than about $5 \%$ in the case of the ideal structure, we consider only the spin magnetization as the main contribution to the total magnetization. This quantity is directly available from spin density functional theory and thus easier to obtain. Furthermore, the small contribution of the orbital moment can be neglected, if we expect a sensitivity of the magnetic properties to structural changes, which should affect the magnetization profiles by more than a few percent.

We choose Fe$|$ZnSe(001) as an example for the whole class of possible systems, because it is of experimental interest,\cite{Rei00} and the pristine semiconductor surface shows much smaller reconstructions than any III-V semiconductor.\cite{Oht99,Mir99} A II-VI material as ZnSe can satisfy the well established but not rigorous counting rules\cite{Har79,Mir99} by building simple  dimers at the (001) surface. Hence, the often used GaAs(001) surface is much more complicated.

% Method

\section{Method}
\label{sec:Method}

Density functional theory (DFT)\cite{Hoh64,Koh65} provides a reliable \textit{ab initio} method for computing the ground-state energy and the associated one-particle density of an electronic many-particle system with given external potential. Thus, it is possible to determine the ground-state configuration of ions in a chemical system, such as molecules or crystals. Due to the ferromagnetic character of iron we need to consider the spin degree of freedom. This possibility is offered by spin density functional theory (SDFT).\cite{Bar72,Pan72,Raj73} Within this generalization of the original DFT the ground-state energy is a functional of the density as well as of the magnetization.

\begin{table*}
\begin{ruledtabular}
\begin{tabular}{lccccccccc}
& 
\multicolumn{3}{c}{$a\;[\mbox{\AA}]$} &
\multicolumn{3}{c}{$K_0\;[\mbox{GPa}]$} &
\multicolumn{3}{c}{$K_0^\prime$} \\
crystal & LDA & GGA & Exp & LDA & GGA & Exp & LDA & GGA & Exp\\
\hline
$\alpha$-Fe & 
$2.761$ & $2.864$ & $2.8662$ & 
$232$ & $152$ & $167$ &
$5.04$ & $5.06$ & $5.97$\\
$\gamma$-Fe & 
$3.387$ & $3.530$ & $3.6468$ & 
$282$ & $155$ & --- &
$6.00$ & $5.33$ & ---\\
ZnSe & 
$5.578$ & $5.741$ & $5.6686$ & 
$68.9$ & $55.5$ & $62.4$ &
$4.73$ & $4.30$ & $4.77$
\end{tabular} 
\end{ruledtabular}
\caption{Cubic lattice constants of iron and zincselenide as results of LDA and GGA calculations, bulk moduli $K_0$ and the according pressure derivatives $K_0^\prime$. The experimental data is taken from Refs.\ \onlinecite{LBIII6,LBIII11,Lee73}. The lattice constant for $\gamma$-Fe is measured at a temperature of $916\;^\circ\mathrm C$, the other data corresponds to room temperature.
\label{tab:lattice_constants}
}
\end{table*}

In this work, we use the \textit{ab initio} package VASP\cite{Kre93,Kre96,VASP} for the numerical implementation of DFT and SDFT. Due to the three-dimensional lattice symmetry of the bulk systems the wave functions of the valence electrons are expanded in in terms of plane waves. The core states are accounted for by ultrasoft pseudopotentials\cite{Kre94} similar to Vanderbilt's potentials,\cite{Van90} which are no longer norm conserving. The exchange and correlation energy is treated within the local-density approximation (LDA) and the generalized-gradient approximation (GGA). In all LDA functionals the parameterization of Perdew and Zunger\cite{Per81} to Monte-Carlo data of Ceperly and Alder\cite{Cep80} is used; in the GGA functionals the proposal of Perdew and Wang (1991)\cite{Per92} is adopted.

Using these techniques we are able to determine the minimal-energy lattice constants of zincblende ZnSe, body-centered cubic (bcc) $\alpha$-Fe, and face-centered cubic (fcc) $\gamma$-Fe. Therefore, we compute the energy per unit cell as a function of the cubic lattice constant. The cut-off energies for the expansion in in terms of plane waves are $209.5\:\mbox{eV}$ for ZnSe and $297\:\mbox{eV}$ for Fe. The momentum-space integrations are done by summing over a grid of special points according to Monkhorst and Pack.\cite{Mon76} Here we use a $10\times10\times10$ mesh for ZnSe with two atoms per unit cell and $20\times20\times20$ meshes for both modifications of iron each with one atom per cell. The results of the Murnaghan\cite{Mur44} fits to the results are given in Table \ref{tab:lattice_constants}. The lattice constants agree very well with the experimental data and with the result of former calculations.\cite{Zhu92,Hag93,Lee95,Asa99} Also the bulk moduli and their pressure derivatives, which result from the Murnaghan fits, show good agreement with experiment and computation. Additionally, iron is ferromagnetic in the bcc structure. The spontaneous magnetic moment per atom is $\mu=2.04\;\mu_{\mathrm B}$ using LDA and $\mu=2.35\;\mu_{\mathrm B}$ in a GGA computation. Thus, the values $\mu=2.12\;\mu_{\mathrm B}$ using LDA and $\mu=2.35\;\mu_{\mathrm B}$ using GGA of a former calculation\cite{Zhu92} can be reproduced. These results soundly fit to the experimental\cite{LBIII19a} value of $\mu=2.23\;\mu_{\mathrm B}$. However, using LDA $\gamma$-Fe turns out to be stable in fcc and not the actual bcc structure. This inconsistence with the experiment is an artifact of the local-density approximation\cite{Bag83} which can be repaired using generalized-gradient corrections. 

Surfaces are treated within the periodic-slab method.\cite{Che76} The program package VASP needs three-dimensional lattice symmetry, which is broken in the case of a surface. But it is possible to restore periodicity perpendicular to the surface by periodically stacked slabs of material. Here we take into account nine atomic layers of the semiconductor ZnSe covered by one, two, three, or four layers of iron. The dangling bonds at the uncovered side of ZnSe are saturated with hydrogen atoms. The distance between the slabs is about $8.5\:\mbox{\AA}$. Tests showed that these parameters are a good compromise between computational effort and accuracy. More details about the geometry of the used supercell can be taken from Sec.\ \ref{sec:ideal} and Fig.\ \ref{fig:ideal}. In these computations we use a $6\times6\times1$ mesh of special points and an energy cut-off of $237.5\:\mbox{eV}$. The ionic relaxation to the equilibrium structure is determined using a conjugate-gradient algorithm.\cite{Pay92}

\section{The ideal interface}
\label{sec:ideal}

The lattice mismatch $\gamma$ of $\alpha$-Fe and ZnSe can be defined as follows:
\begin{equation}
\gamma = \frac{a_{\mathrm{ZnSe}}-2a_{\mathrm{Fe}}} {\frac12\left(a_{\mathrm{ZnSe}}+2a_{\mathrm{Fe}}\right)}\;.
\end{equation}
Comparison with Table \ref{tab:lattice_constants} shows that this mismatch is $\gamma= 1.0 \%$ using LDA, $\gamma= 0.2 \%$ using GGA, and $\gamma= -1.1 \%$ for the experimental values. Thus, the lattice mismatch is very small, and the idea of an ideal structure of the interface is suggestive. For many systems of iron grown on the (001) surfaces of zincblende semiconductors the mismatch of the lattice constants is quite small (experimental values:\cite{LBIII6} e.g.,\ $a_{\rm GaAs}=5.653\:\mbox{\AA}$,  $a_{\rm AlAs}=5.660\:\mbox{\AA}$). Hence, many calculations\cite{But97,Mac98,Kos00} are based on this idealization.

\begin{figure}[b]
  \psfrag{aF}{\raisebox{0.2ex}{\hspace{0.4ex}$a_{\mathrm{Fe}}$}}
  \psfrag{aZS}{\raisebox{0.4ex}{$a_{\mathrm{ZnSe}}$}}
  \psfrag{= Fe}{$\equiv$\hspace{1ex} Fe}
  \psfrag{= Zn}{$\equiv$\hspace{1ex} Zn}
  \psfrag{= Se}{$\equiv$\hspace{1ex} Se}
  \psfrag{= H}{$\equiv$\hspace{1ex} H}
  \includegraphics[width=\factorgr\linewidth]{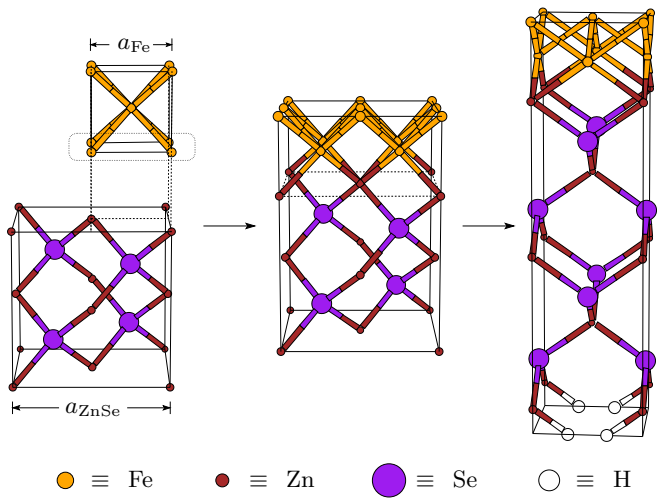}
  \caption
  {
    Structure of the ideal interface. Here the semiconductor surface
    is Zn-terminated. 
    4 Fe cubes fit on a single ZnSe cube as displayed on the left. Thus, the
    structure shown in the middle is plausible. 
    The elementary cell used for the calculation is presented on the
    right.
  \label{fig:ideal}
  }
\end{figure}

Fig.~\ref{fig:ideal} gives an intuitive picture of the ideal interface. Regarding the pristine surface of ZnSe first, it is evident that (001) surfaces always are terminated by only one sort of atom, thus we have to distinguish between the Se- and Zn-terminated case. Due to the tiny lattice mismatch four iron cubes closely fit on a single ZnSe cube. Substitution of the lowermost atomic layer of iron by the first semiconductor layer results in the ideal structure shown in the middle of Fig.~\ref{fig:ideal}. In this case the semiconductor is Zn-terminated. 

But the presented cell is not elementary in a two-dimensional sense parallel to the surface. However, an elementary cell can be obtained by rotating the structure around an axis perpendicular to the surface by an angle of $45^\circ$ and cutting off the corners. This is visualized on the right side of Fig.~\ref{fig:ideal} by showing the elementary cell used in the calculations for the Zn-terminated semiconductor covered by two monolayers of iron. As already mentioned in Sec.\ \ref{sec:Method} we choose 9 atomic layers of ZnSe and 1, 2, 3, or 4 monolayers of iron. The separating vacuum slab of a thickness of about $8.5\:\mbox{\AA}$ is not displayed. A monolayer in ZnSe contains either one Zn or one Se atom in the cell, a single monolayer of Fe contains two inequivalent atoms. The dangling bonds at the bottom side are saturated with hydrogen atoms. The positions of the H atoms and the lowermost atoms of ZnSe are fixed in the calculations; thus, the bond distances between H and Se or Zn are constant with $d_{\mathrm{HSe}}=1.478\:\mbox\AA$ and $d_{\mathrm{HZn}}=1.612\:\mbox\AA$. These values are the equilibrium bond distances of the corresponding isolated dimers computed with VASP. 

Starting from the ideal structure it is possible to compute the relaxation of the atoms in the elementary cell. If restricted to the direction parallel to the surface normal the displacements turn out to be small, and thus the structure of the surface remains nearly ideal. However, this constraint may lead to possibly unstable equilibrium positions. In order to allow relaxation into reconstructed equilibrium positions we start from randomly displaced positions of the peripheral Fe atoms. The results of these computations may differ dramatically from the ideal structure. This is reported in the following section.

\section{Results}

In this section we present the results obtained for Se- and Zn-terminated (001) surfaces of ZnSe covered by 1, 2, 3, and 4 monolayers of iron. In all cases the energy-minimizing structure is determined first. For an estimation of the dependence of magnetic properties on the structure we show the profiles of the spin magnetization perpendicular to the interface. The term \emph{profile} means that the corresponding magnitude is averaged over planes parallel to the surface. In bulk iron the orbital magnetization is suppressed because of quenching by electric crystal fields.\cite{Bra66} At the surface or interface the crystal symmetry is broken; thus, the quenching does not suffice any more to disable completely the orbital contribution to the magnetization. Nevertheless, in calculations\cite{Kos00} considering the orbital contribution this appears to be not more than a $6 \%$ effect. Thus, the mere spin polarization is a good indicator for the behavior of the magnetic properties when structural deviations from the ideal case have to be expected.

\subsection{A single monolayer of iron}

\begin{figure}[b]
\begin{center}
  \psfrag{d1}{$d_1$}
  \psfrag{d2}{$d_2$}
  \psfrag{d3}{$d_3$}
  \psfrag{d4}{$d_4$}
  \psfrag{d5}{$d_5$}
  \psfrag{d6}{$d_6$}
  \psfrag{a}{$\alpha$}
  \psfrag{b}{$\beta$}
  \psfrag{g}{$\gamma$}
  \psfrag{Se-terminated}{Se-terminated}
  \psfrag{Zn-terminated}{Zn-terminated}
  \includegraphics[width=\factorkl\linewidth]{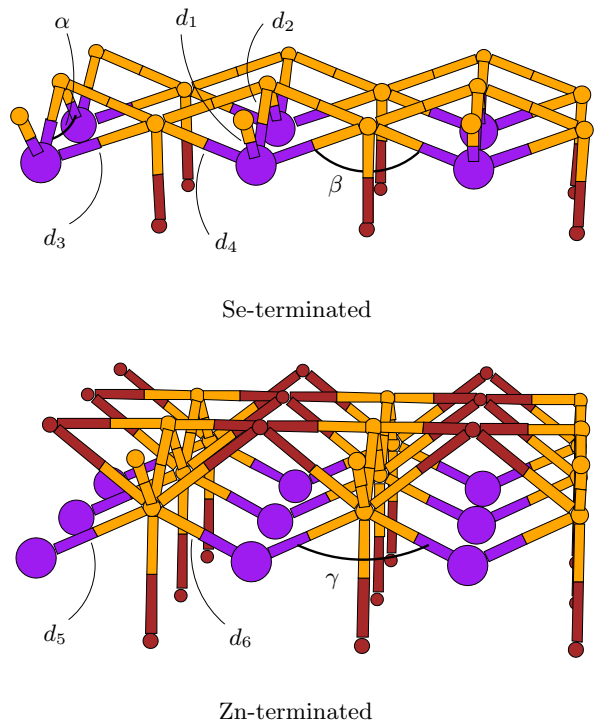}
\end{center}
  \caption{
	Structure of the Se-terminated (upper part) and Zn-terminated (lower 
	part) ZnSe(001) surface covered by a single monolayer of iron. Only the
	crucial region near the interface with interesting distances and angles 
	is shown.
  \label{fig:1ML_zoom}
  }
\end{figure}

The most interesting results are obtained for only one Fe monolayer on ZnSe(001). In particular, the relaxed structure deviates tremendously from the ideal case. Fig.~\ref{fig:1ML_zoom} shows the peripheric region of the covered Se- and Zn-terminated surface. In the Se-terminated case the two inequivalent Fe atoms obviously have very different vertical positions. One Fe atom is lowered, which enables a bond to a Zn atom of the second atomic layer of ZnSe. The other Fe atom is bonded to the uppermost Se and the other Fe atom. Fe and Se together build up a surface slab that is coupled to the ZnSe substrate only by an Fe--Zn bond. So there is no bond between Se and Zn at the two uppermost layers any more. In this slab, the the Fe and Se ions form zigzag chains from the front backwards (in Fig.~\ref{fig:1ML_zoom}) with bond lengths 
$d_1=2.30\:\mbox\AA$, $d_2=2.42\:\mbox\AA$, and Se--Fe--Se bond angle $\alpha=119^\circ$. Similar chains run from the left to the right with the values $d_3=2.34\:\mbox\AA$, $d_4=2.33\:\mbox\AA$, and $\beta=121^\circ$. 

The Zn-terminated surface provides the most amazing results. The uppermost Zn layer is buoying upwards which can be seen at the lower part of Fig.~\ref{fig:1ML_zoom}. We obtain two layers in the perpheric region. The uppermost layer consists of one Fe and one Zn atom in the elementary cell with nearly identical heights. The second layer is made up of one Fe and one Se atom per cell. Similar to the Se-terminated case these atoms build zigzag chains sidewards. There the bond distances are
$d_5=2.44\:\mbox\AA$, $d_6=2.42\:\mbox\AA$, and the Se--Fe--Se angle is $\beta=121^\circ$. Again, the uppermost Se atom is only bonded to iron, and the hole peripheric layer is coupled to the semiconductor by an Fe--Zn bond. Both structural results can be interpreted in the way that non-metallic Se forms tight bonds to Fe. This affinity is so distinctive that iron can even displace the originally terminating Zn atoms towards the outermost region.

\begin{figure}
  \psfrag{ideal}{\scriptsize ideal}
  \psfrag{relax.}{\scriptsize relax.}
  \psfrag{z [A]}{\raisebox{-0.9ex}{$z\;[\mbox{{\AA}}]$}}
  \psfrag{n(z) [A-3]}{\raisebox{-0.4ex}{$\overline n(z)\;[\mbox{{\AA}}^{-3}]$}}
  \psfrag{m(z) [mb A-3]}{\raisebox{-0.4ex}{$\overline m_z(z)\;[\mathrm{\mu_B\mbox{{\AA}}^{-3}}]$}}
  \includegraphics[width=\factorgr\linewidth]{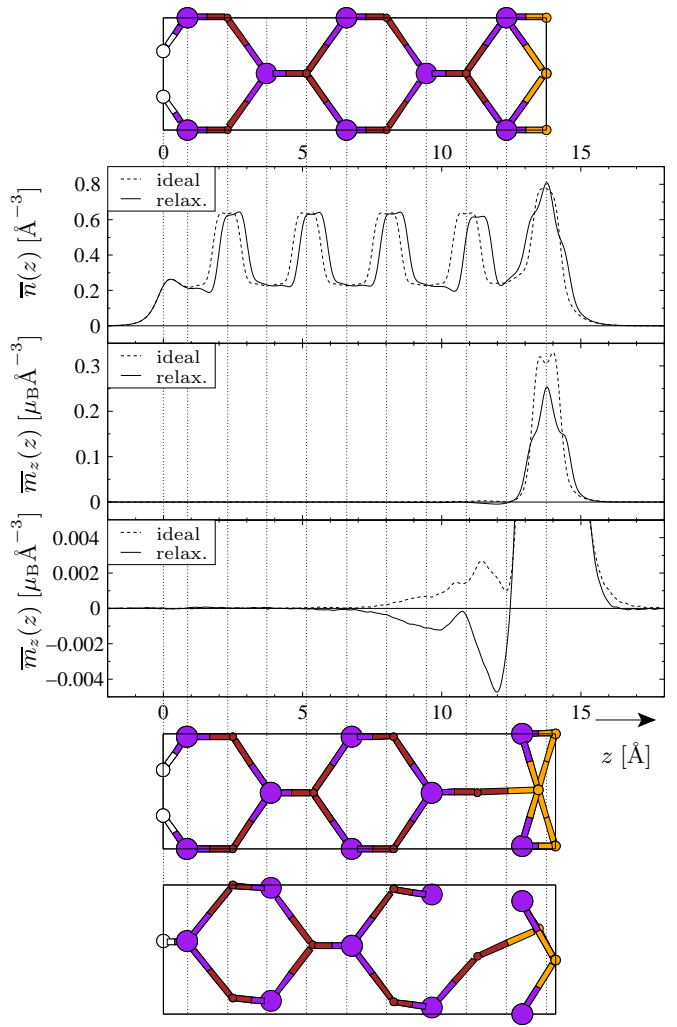}
  \caption{
Structure and density profiles for one monolayer of Fe (2 atoms per cell) on Se-terminated ZnSe(001). The vertical dotted lines mark the positions of the atomic layers in the ideal case. At the top the elementary cell of the ideal interface is presented. At the bottom the atoms are shown after relaxation in two different perspectives. The uppermost plot shows the density of the valence electrons, in the middle the magnetization is displayed, and in the lowermost plot the magnetization is shown on an expanded scale. The solid (dashed) lines belong to the relaxed (ideal) structure. 
  \label{fig:1ML_Se_prof}
  }
\end{figure}

An indicator for the stability of the surface structure is the energy gain due to the relaxation in comparison with the ideal structure. Also the results for DFT and SDFT have to be differentiated. Because the relaxation was done initially using usual DFT the result of which was refined with SDFT afterwards, we have to distinguish between energies from DFT and SDFT. For the Se-terminated surface the energy gain is $\Delta E_\perp=0.638\:\mbox{eV}$ from the small relaxation of the peripheric atom perpendicular to the interface. Complete DFT relaxation causes an increase to $\Delta E_{\rm DFT}=1.663\:\mbox{eV}$ by more than $1\:\mbox{eV}$. This value is much smaller using SDFT even after further relaxations. The smaller energy gain $\Delta E_{\rm SDFT}=0.346\:\mbox{eV}$ can be interpreted in the following way: DFT only can give the energy differences due to structural changes, whereas using SDFT this is corrected by the energy due to the ordering of the spins. Thus, the ferromagnetic order in the ideal structure is energetically advantageous in comparison with the relaxed structure. Nevertheless, the structural contribution to $\Delta E$ outbalances the ferromagnetic one; thus, the energy gain is still much greater than the thermal energy $k_{\rm B}T\approx 25\:\mbox{meV}$ corresponding to  room temperature. 

\begin{figure}
  \psfrag{ideal}{\scriptsize ideal}
  \psfrag{relax.}{\scriptsize relax.}
  \psfrag{z [A]}{\raisebox{-0.9ex}{$z\;[\mbox{\AA}]$}}
  \psfrag{n(z) [A-3]}{\raisebox{-0.4ex}{$\overline n(z)\;[\mbox{{\AA}}^{-3}]$}}
  \psfrag{m(z) [mb A-3]}{\raisebox{-0.4ex}{$\overline m_z(z)\;[\mathrm{\mu_B\mbox{{\AA}}^{-3}}]$}}
  \includegraphics[width=\factorgr\linewidth]{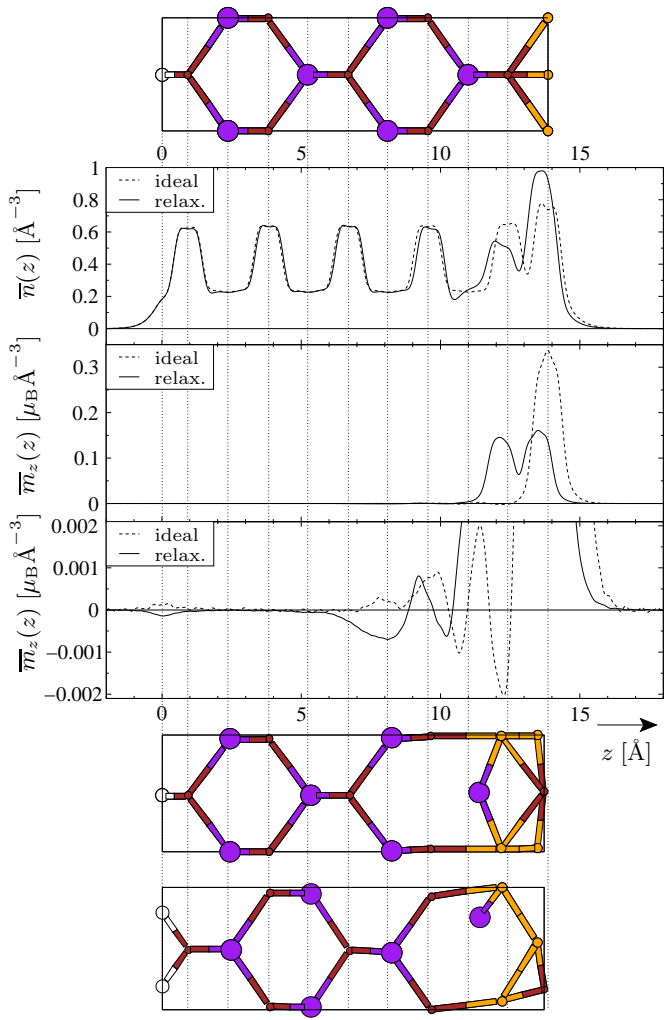}
  \caption{
	Same as Fig.~\ref{fig:1ML_Se_prof}, but for one monolayer of Fe on
	Zn-terminated ZnSe(001).
  \label{fig:1ML_Zn_prof}
  }
\end{figure}

In the case of the Zn-terminated surface the respective values are $\Delta E_\perp=0.125\:\mbox{eV}$, $\Delta E_{\rm DFT}=2.767\:\mbox{eV}$, and $\Delta E_{\rm SDFT}=0.577\:\mbox{eV}$. $\Delta E_\perp$ is smaller than in the Se-terminated case, because the reduction of the distance between the metal Fe and the nonmetal Se yields a higher energy gain than a variation of the Fe--Zn bond length. In the DFT calculation, $\Delta E_{\rm DFT}$ is much larger than in the Se-terminated case. So the energetic advantage of the Zn atoms buoying upwards is enormous. On the other hand, this dramatic structural variation reduces the energy gain due to ferromagnetism more than for the Se-terminated surface. But still $\Delta E_{\rm SDFT}$ is much greater than typical thermal energies at room temperature.

The density profiles measure the dependence of electronic and magnetic properties on structural variations. The profiles for the density of the valence electrons, the magnetization, and the magnetization on an expanded scale are shown in Figs.~\ref{fig:1ML_Se_prof} and \ref{fig:1ML_Zn_prof}. Additionally, the atomic positions in the elementary cells are presented for the ideal and relaxed structure. The valence-charge density has minima for Se layers and maxima for Zn and Fe layers. This is because Se is treated with a pseudopotential for 6 valence electrons ($4s^24p^4$), whereas for Zn 12 electrons ($3d^{10}4s^2$) are taken into account, although the total number of electrons is larger for Se ($Z=34$) than that of Zn ($Z=30$). All the changes of the electron density can be interpreted as a shift of the density along the direction of the corresponding relaxations. Also the magnetization seems to follow the ionic relaxations, especially for the Fe layers, although the total magnetic moment per cell decreases from $6.32\:\mu_{\rm B}$ for the ideal interface to $4.94\:\mu_{\rm B}$ after relaxation in the Se-terminated case and from $6.17\:\mu_{\rm B}$ to $5.34\:\mu_{\rm B}$, if the ZnSe surface was Zn-terminated. The exact run of the curves in Figs.~\ref{fig:1ML_Se_prof} and \ref{fig:1ML_Zn_prof} according to the relaxed structure differs from the ideal case, which is clear because of the splitting of the Fe plane into two layers.

The lowermost graphics in both, Fig.~\ref{fig:1ML_Se_prof} and  \ref{fig:1ML_Zn_prof}, show drastic differences between the magnetization before and after relaxation on an expanded scale. In the case of the Se-terminated surface (Fig.~\ref{fig:1ML_Se_prof}) the magnetization inside the semiconductor even has the direction opposite to the total moment. This antiferromagnetic ordering is obviously an effect of the structural changes, because for the ideal interface the magnetization has the same direction inside the semiconductive and the ferromagnetic region. An appropriate interpretation is given by the Heisenberg model.\cite{Mad78} There, the coupling constants $J_{ij}$ are given by the overlap of the wave functions of neighboring atoms. Because the comparatively strongly localized $d$-electrons dominate the ferromagnetism of Fe, a small change of bond lengths or layer distances can invert the sign of the coupling constants, which means a transition from ferro- to antiferromagnetism. Additionally, the relaxation results in an Fe--Zn bond that does not exist in the ideal case. In contrast to Fe--Se bonds, these bonds lead to a flip of the magnetization, as can be seen for the Zn-terminated interface. For the Zn-terminated surface (Fig.~\ref{fig:1ML_Se_prof}) a more complicated run of the curves inside ZnSe is obtained. Fe--Zn bonds at the interface instead of Fe--Se bonds for Se termination cause antiferromagnetic effects, even for the ideal structure. However, the magnetization deep inside the semiconductor has flipped its direction after relaxation, too. The tiny peaks at the position of the hydrogen atoms indicate too thin slabs or vacuum layers. But incrementing those two parameters would require a drastic increase of computing time. Because we just want to estimate the dependence of magnetic properties on the structure, the chosen accuracy is sufficient.

\subsection{Two, three, and four monolayers}

Now we turn to the case of 2, 3, or 4 monolayers of Fe grown on ZnSe(001). The structure and the associated density profiles are presented in Figs.\ \ref{fig:2ML_Se_prof}, \ref{fig:3ML_Se_prof}, and \ref{fig:4ML_Se_prof} for the Se-terminated surfaces and in Figs.\ \ref{fig:2ML_Zn_prof}, \ref{fig:3ML_Zn_prof}, and \ref{fig:4ML_Zn_prof} for the Zn-terminated case. Here, we omit the presentation of the structure like the one in Fig.~\ref{fig:1ML_zoom} in a separate plot. In the results for a given termination clear similarities arise, thus a common discussion is sensible.

\subsubsection{Se termination}

\begin{figure}[b]
  \psfrag{ideal}{\scriptsize ideal}
  \psfrag{relax.}{\scriptsize relax.}
  \psfrag{z [A]}{\raisebox{-0.9ex}{$z\;[\mbox{\AA}]$}}
  \psfrag{n(z) [A-3]}{\raisebox{-0.4ex}{$\overline n(z)\;[\mbox{{\AA}}^{-3}]$}}
  \psfrag{m(z) [mb A-3]}{\raisebox{-0.4ex}{$\overline m_z(z)\;[\mathrm{\mu_B\mbox{{\AA}}^{-3}}]$}}
  \includegraphics[width=\factorgr\linewidth]{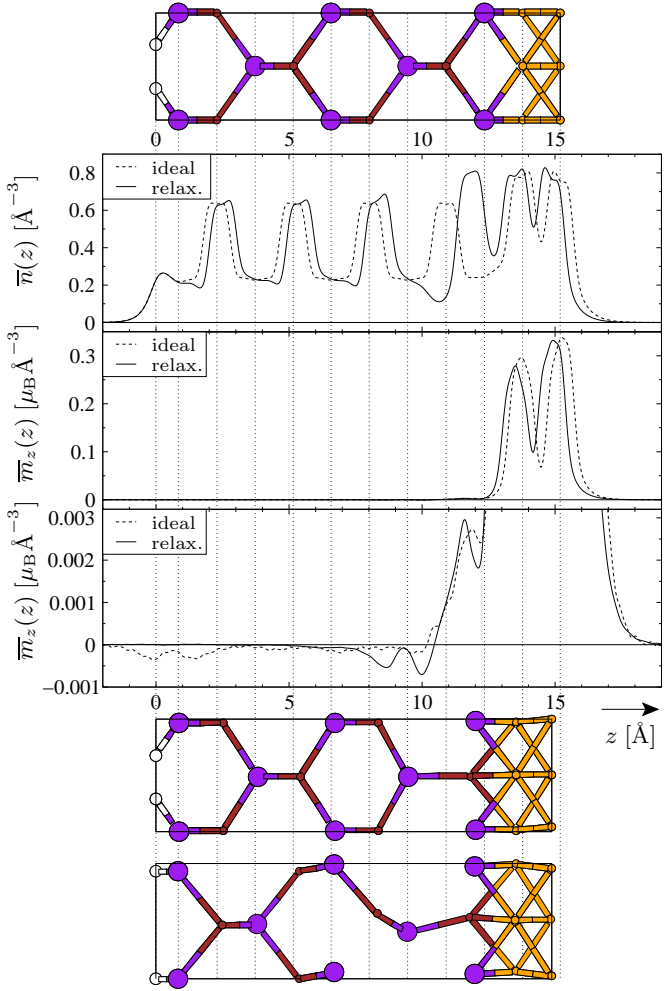}
  \caption
  {
	Same as Fig.~\ref{fig:1ML_Se_prof}, but 
	for two monolayers of Fe on Se-terminated ZnSe(001).
  \label{fig:2ML_Se_prof}
  }
\end{figure}

\begin{figure}
  \psfrag{ideal}{\scriptsize ideal}
  \psfrag{relax.}{\scriptsize relax.}
  \psfrag{z [A]}{\raisebox{-0.9ex}{$z\;[\mbox{\AA}]$}}
  \psfrag{n(z) [A-3]}{\raisebox{-0.4ex}{$\overline n(z)\;[\mbox{{\AA}}^{-3}]$}}
  \psfrag{m(z) [mb A-3]}{\raisebox{-0.4ex}{$\overline m_z(z)\;[\mathrm{\mu_B\mbox{{\AA}}^{-3}}]$}}
  \includegraphics[width=\factorgr\linewidth]{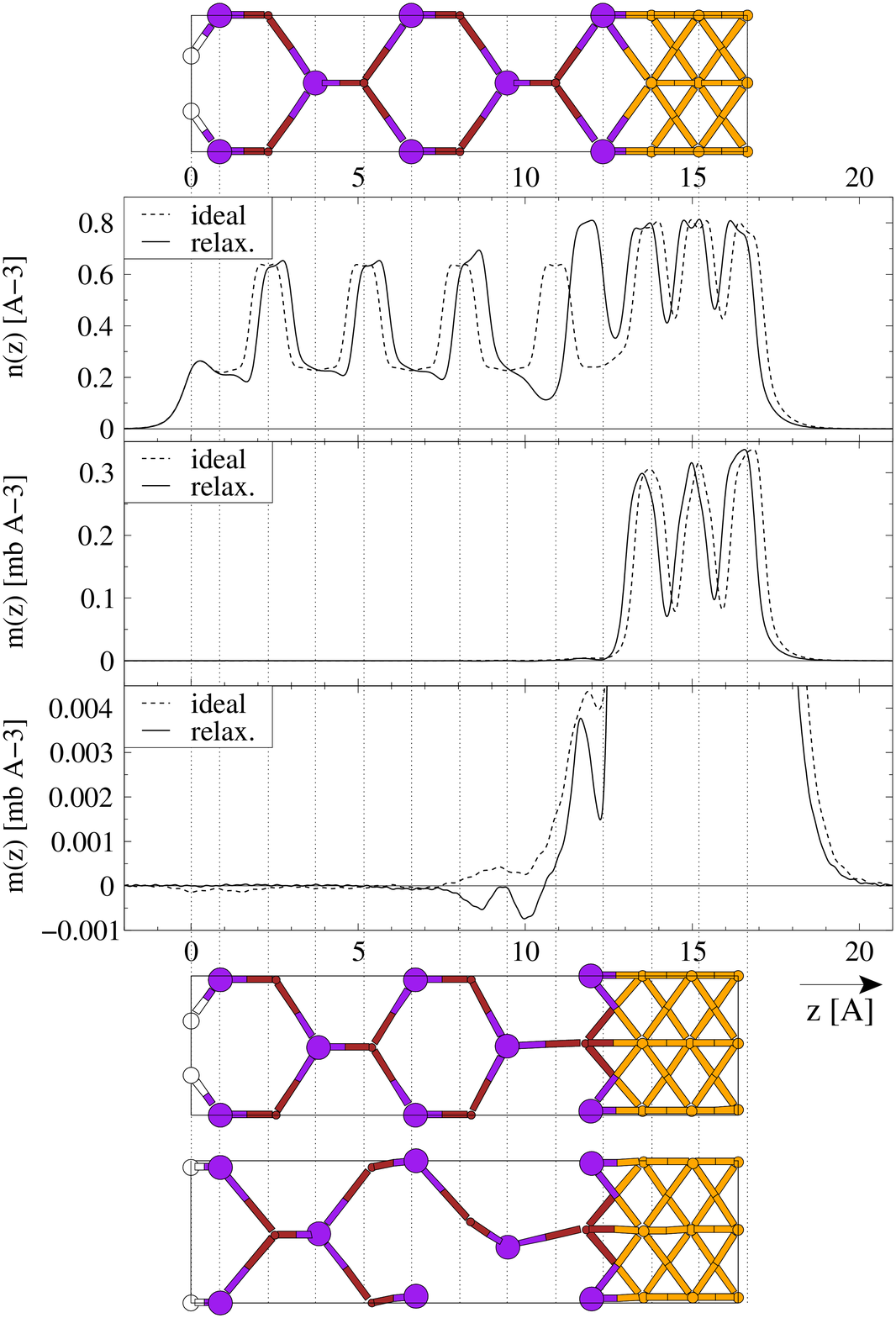}
  \caption
  {
	Same as Fig.~\ref{fig:1ML_Se_prof}, but
	for three monolayers of Fe on Se-terminated ZnSe(001).
  \label{fig:3ML_Se_prof}
  }
\end{figure}

\begin{figure}
  \psfrag{ideal}{\scriptsize ideal}
  \psfrag{relax.}{\scriptsize relax.}
  \psfrag{z [A]}{\raisebox{-0.9ex}{$z\;[\mbox{\AA}]$}}
  \psfrag{n(z) [A-3]}{\raisebox{-0.4ex}{$\overline n(z)\;[\mbox{{\AA}}^{-3}]$}}
  \psfrag{m(z) [mb A-3]}{\raisebox{-0.4ex}{$\overline m_z(z)\;[\mathrm{\mu_B\mbox{{\AA}}^{-3}}]$}}
  \includegraphics[width=\factorgr\linewidth]{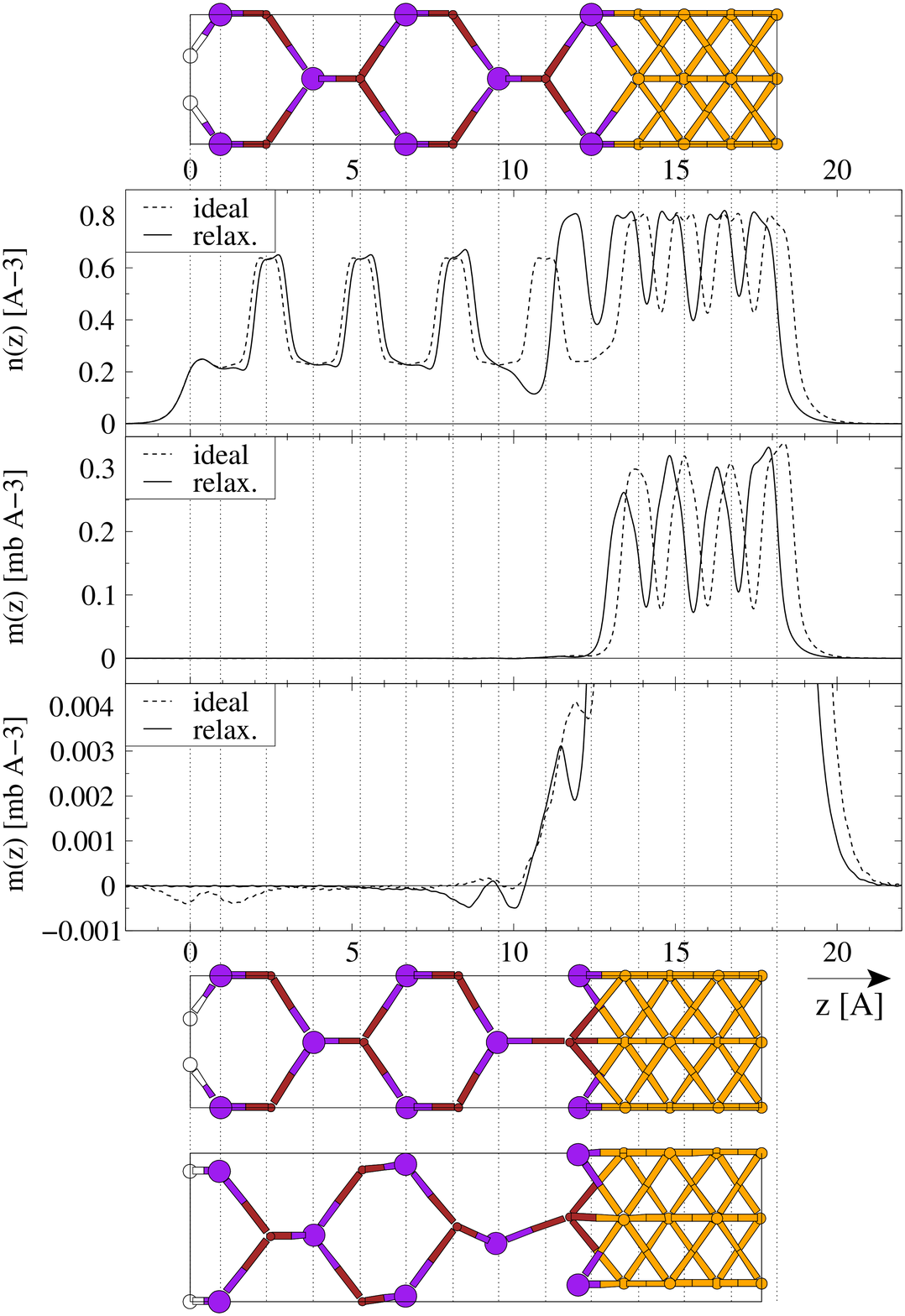}
  \caption
  {
	Same as Fig.~\ref{fig:1ML_Se_prof}, but
	for four monolayers Fe on Se-terminated ZnSe(001).
  \label{fig:4ML_Se_prof}
  }
\end{figure}

The structures of all three systems are very similar. Those are presented beneath the plots of the profiles in Figs.\ \ref{fig:2ML_Se_prof} (2 monolayers), \ref{fig:3ML_Se_prof} (3 monolayers), and \ref{fig:4ML_Se_prof} (4 monolayers). The iron slab has nearly perfect bcc structure as known from bulk $\alpha$-Fe. Relative relaxations parallel and perpendicular to the interface are negligibly small. However, the whole Fe region is relaxing towards the semiconductor substrate by several $0.1\:\mbox{\AA}$, which is an effect of structural changes in ZnSe. The first atomic layer of ZnSe does not solely exist of Se any more. The Zn atoms of the originally second atomic layer together with the terminating Se atoms build a ZnSe layer at the interface. So it is possible that the uppermost Se atoms are not bonded to Zn but to Fe, thus the whole surface slab is coupled to the substrate by Fe--Zn bonds followed by a Zn--Se bond. The distance of the second Se layer, which can be regarded as the uppermost layer of the substrate, to the first layer of the peripheric part of the slab (the ZnSe layer) is clearly greater than the layer distances in the ideal structure. So the peripheric part of the slab should be coupled weakly to the substrate, and it can be displaced easily in the horizontal direction. This would implicate a very soft phonon mode. Measuring such a mode could confirm our results for the structure. Deeper in the semiconductor the structure becomes more and more similar to that of zincblende. Beyond the fourth atomic layer (second Zn layer) the structure can be regarded as ideal. The energy gain per cell decreases with increasing number of Fe layers from $0.451\:\mbox{eV}$ to $0.392\:\mbox{eV}$ and $0.279\:\mbox{eV}$ due to the increasingly ideal character of the ionic configuration.

The almost ideal structure of the iron is reflected in the nearly ideal profiles of the valence electron density and the magnetization in the Fe region, too. The peaks corresponding to iron are simply shifted along the direction of the ionic relaxation. Only the peak nearest to the interface is slightly decreased for the case of two and four monolayers of Fe. So the total magnetic moment per cell reduces from $11.14\:\mu_{\rm B}$ for the ideal structure with two Fe layers to $10.68\:\mu_{\rm B}$ after relaxation, from $16.79\:\mu_{\rm B}$ to $16.35\:\mu_{\rm B}$ for three monolayers and from $21.71\:\mu_{\rm B}$ to $20.49\:\mu_{\rm B}$ for four monolayers. The ZnSe layer at the interface entails an additional peak in the electronic density, because this layer is packed very densely. Towards ZnSe this peak is followed by a dip, which clearly shows the already assumed weakness of the expanded Zn--Se bond. Inside the semiconductor the density corresponds to a shift along the relaxation again. The magnetization inside ZnSe declines very fast in all three cases. However, the direction of the weak magnetization is flipped, similar to the case of a single monolayer of Fe (compare Fig.\ \ref{fig:1ML_Se_prof}). Again different layer distances and changing chemical environments result in antiferromagnetic ordering. In all three cases the magnitude of the magnetization and the penetration depth slightly increase.

\subsubsection{Zn termination}

\begin{figure}[b]
  \psfrag{ideal}{\scriptsize ideal}
  \psfrag{relax.}{\scriptsize relax.}
  \psfrag{z [A]}{\raisebox{-0.9ex}{$z\;[\mbox{\AA}]$}}
  \psfrag{n(z) [A-3]}{\raisebox{-0.4ex}{$\overline n(z)\;[\mbox{{\AA}}^{-3}]$}}
  \psfrag{m(z) [mb A-3]}{\raisebox{-0.4ex}{$\overline m_z(z)\;[\mathrm{\mu_B\mbox{{\AA}}^{-3}}]$}}
  \includegraphics[width=\factorgr\linewidth]{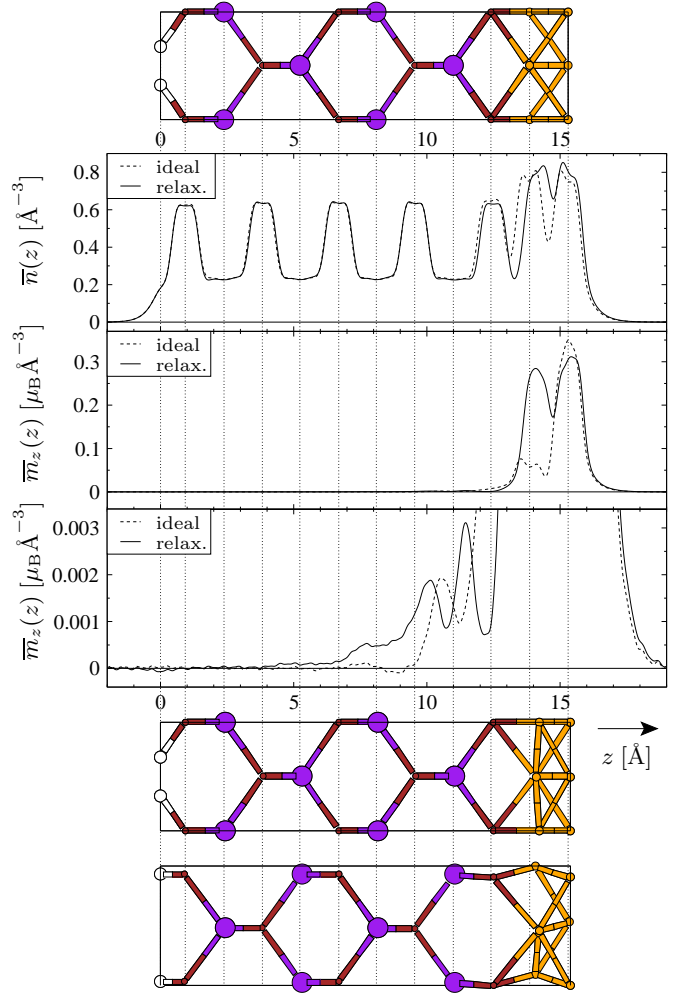}
  \caption
  {
	Same as  Fig.~\ref{fig:1ML_Se_prof}, but 
	for two monolayers of Fe on Zn-terminated ZnSe(001).
  \label{fig:2ML_Zn_prof}
  }
\end{figure}

\begin{figure}
  \psfrag{ideal}{\scriptsize ideal}
  \psfrag{relax.}{\scriptsize relax.}
  \psfrag{z [A]}{\raisebox{-0.9ex}{$z\;[\mbox{\AA}]$}}
  \psfrag{n(z) [A-3]}{\raisebox{-0.4ex}{$\overline n(z)\;[\mbox{{\AA}}^{-3}]$}}
  \psfrag{m(z) [mb A-3]}{\raisebox{-0.4ex}{$\overline m_z(z)\;[\mathrm{\mu_B\mbox{{\AA}}^{-3}}]$}}
  \includegraphics[width=\factorgr\linewidth]{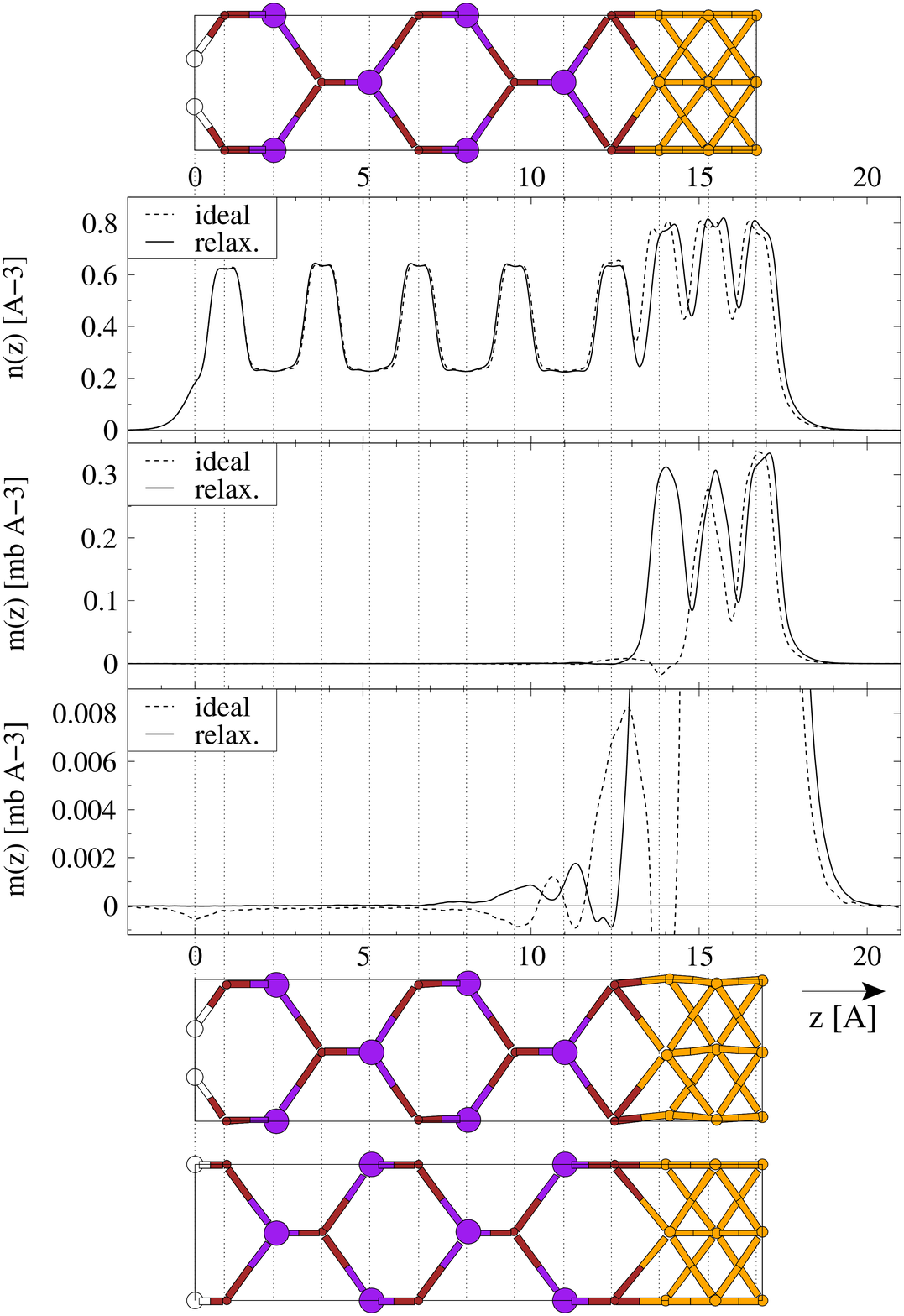}
  \caption
  {
	Same as  Fig.~\ref{fig:1ML_Se_prof}, but
	for three monolayers of Fe on Zn-terminated ZnSe(001).
  \label{fig:3ML_Zn_prof}
  }
\end{figure}

\begin{figure}
  \psfrag{ideal}{\scriptsize ideal}
  \psfrag{relax.}{\scriptsize relax.}
  \psfrag{z [A]}{\raisebox{-0.9ex}{$z\;[\mbox{\AA}]$}}
  \psfrag{n(z) [A-3]}{\raisebox{-0.4ex}{$\overline n(z)\;[\mbox{{\AA}}^{-3}]$}}
  \psfrag{m(z) [mb A-3]}{\raisebox{-0.4ex}{$\overline m_z(z)\;[\mathrm{\mu_B\mbox{{\AA}}^{-3}}]$}}
  \includegraphics[width=\factorgr\linewidth]{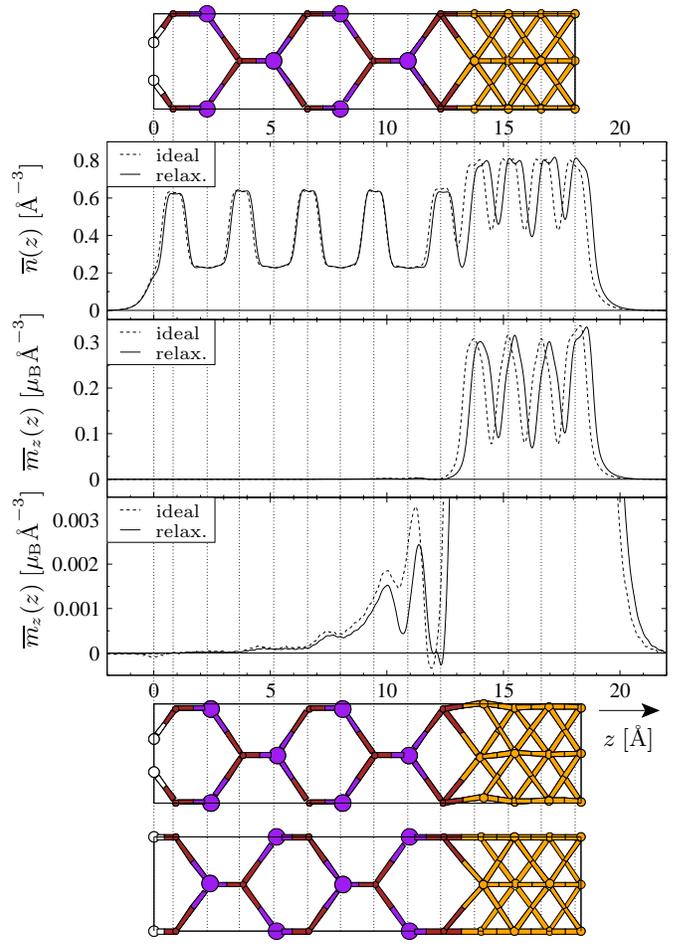}
  \caption
  {
	Same as Fig.~\ref{fig:1ML_Se_prof}, but
	for four monolayers of Fe on Zn-terminated ZnSe(001).
  \label{fig:4ML_Zn_prof}
  }
\end{figure}

For the Zn-terminated surfaces a different picture of the structure is obtained. Here the deviations from ideal behavior preferentially appear in the Fe slab, whereas ZnSe remains in the almost perfect zincblende structure. Inspection of Figs.\ \ref{fig:2ML_Zn_prof}, \ref{fig:3ML_Zn_prof}, and \ref{fig:4ML_Zn_prof} demonstrates that the relaxations in the semiconductor region are nearly vanishing. There is no drastic structural change at the interface like the formation of a ZnSe layer in the Se-terminated case. However, relaxations in the Fe region are obvious. Here the Fe atoms are displaced in the direction parallel to the interface; thus, the bond lengths change, but the layer distances do not. For only two monolayers of Fe these relaxations are most evident. Actually, the Fe atoms at the interface are able to form additional bonds within the layer. The deviations from the ideal structure seem to become weaker with an increasing number of monolayers. In contrast to the Se-terminated interface the Fe slab slightly relaxes outwards.

Because of the smaller extent of the structural modification in comparison with the Se-terminated case we expect smaller energy gain due to relaxation. But surprisingly the values $1.027\:\mbox{eV}$, $0.814\:\mbox{eV}$, and $0.610\:\mbox{eV}$ are much greater than for Se termination. Changing the ionic configuration in the Fe slab seems to reduce the energy by a greater amount than the relaxations in the semiconductor do. However, the energy gain decreases with the number of Fe layers, which again shows the increasingly ideal character of the system.

The density profiles of valence electrons do not exhibit any astonishing effects. Because there are no closer packed layers or strongly increasing layer distances, only the outwards relaxation of the whole Fe region can be identified in a shift of the Fe peaks. However, the magnetization even within the iron region extremely deviates from the ideal behavior. For the ideal structure the magnetization due to the Fe layer at the interface is strongly suppressed in the cases of 2 (Fig.\  \ref{fig:2ML_Zn_prof}) and 3 (Fig.\ \ref{fig:3ML_Zn_prof}) monolayers of Fe. This result could be interpreted as magnetic dead layers claimed in older work.\cite{Kre87} However, the effect of nearly vanishing magnetization at the interface turns out to be unrealistic: In the case of four monolayers of Fe all magnetic moments are fully developed even for the ideal structure, and in the case of two and three monolayers the dead layers disappear after the relaxation. So we find in agreement with experiment\cite{Zol97} and other theoretical work\cite{Kos00} that there are no magnetic dead layers in all realistic systems considered here. Due to the difficulties in the treatment of the ideal interfaces the total magnetic moment per cell increases for two monolayers from $7.83\:\mu_{\rm B}$ before relaxation to $11.09\:\mu_{\rm B}$ after and for three monolayers from $10.26\:\mu_{\rm B}$ to $16.64\:\mu_{\rm B}$. The known effect of a slight decrease of the moment as seen for Se termination is obtained again for four monolayers of Fe. Here the moment reduces from $21.62\:\mu_{\rm B}$ to $21.52\:\mu_{\rm B}$ by the tiny value of $0.1\:\mu_{\rm B}$. The difference is obviously smaller than for Se termination. Amazingly, the relaxations of the  Fe atoms parallel to the surface can stabilize the magnetic moment, whereas the decrease of the moment is even less pronounced for Se-terminated surfaces, where the iron structure remains almost ideal.

Inside the semiconductor the magnetizations for the ideal and the relaxed structure strongly deviate from each other in the case of two and three monolayers of Fe. This is no surprise because of the dead layers in the ideal structure. So the differences are difficult to interpret, the more so as the lowermost profile in Fig.\ \ref{fig:4ML_Zn_prof} for four monolayers shows only little difference between the two curves corresponding to the ideal and relaxed structure. For all Zn-terminated systems including a single monolayer of Fe a tendency towards oscillations of the magnetization is evident. This is in contrast to Se termination.

\section{Summary and conclusions}
\label{sec:summary}

In this work we have presented structural, electronic, and magnetic results from spin density functional theory calculations for 1, 2, 3, and 4 monolayers of Fe on Se- as well as Zn-terminated ZnSe(001). In spite of the tiny mismatch of the lattice constants of $\alpha$-Fe and cubic ZnSe we find dramatic structural deviations from the ideal interface. Especially for a single monolayer of iron the structure is changed radically. But also for more than one monolayer we obtain amazing results, so we find a separated ZnSe layer with Zn and Se at Fe positions at the interface in the case of Se termination. Otherwise, for Zn termination the structural modifications take place in the Fe region. 

Summarizing, we have to point out that the structures become increasingly ideal with an increasing number of Fe monolayers. We see the tendency of tight Fe--Se bonds; in every Se-terminated system the outermost Se atoms are bonded primarily to Fe. The raised Zn atoms are bonded to Fe and form a single weak bond to the substrate. In the Zn-terminated structures, too, the whole surface layer is coupled to the substrate by an Fe--Zn bond.

We find a clear dependence of the profiles of electronic density and magnetization on the relaxations. The electronic density and the magnetization due to the Fe atoms can be explained by a shift of the curves along the direction of the relaxation. However, the magnetization inside ZnSe often changes the direction in comparison with the ideal structure or the total magnetic moment. Although the magnitude of this magnetization is small, the flip of the sign might affect electronic and magnetic properties near the interface. For example a reliable description of the tunneling magneto resistance (TMR) requires an accurate electronic density of states at the interface.\cite{Mac98,Kos00} However, this magnitude might be sensitive to small relaxations due to changing bond conditions.

Thus, the effect of relaxations should not be neglected in more realistic calculations of electronic and magnetic properties of ferromagnet--semiconductor heterostructures. The usage of the ideal interface in such computations can foil ambitious efforts using complicated methods.

\begin{acknowledgments}
This work was supported by the \emph{Deutsche Forschungsgemeinschaft} with a stipendium in the \emph{Graduiertenkolleg: Komplexit{\"a}t in Festk{\"o}rpern, Phononen, Elektronen und Strukturen}.
\end{acknowledgments}

% Bibliography
%\newpage

\bibliography{cites}

\begin{thebibliography}{42}
\expandafter\ifx\csname natexlab\endcsname\relax\def\natexlab#1{#1}\fi
\expandafter\ifx\csname bibnamefont\endcsname\relax
  \def\bibnamefont#1{#1}\fi
\expandafter\ifx\csname bibfnamefont\endcsname\relax
  \def\bibfnamefont#1{#1}\fi
\expandafter\ifx\csname citenamefont\endcsname\relax
  \def\citenamefont#1{#1}\fi
\expandafter\ifx\csname url\endcsname\relax
  \def\url#1{\texttt{#1}}\fi
\expandafter\ifx\csname urlprefix\endcsname\relax\def\urlprefix{URL }\fi
\providecommand{\bibinfo}[2]{#2}
\providecommand{\eprint}[2][]{\url{#2}}

\bibitem[{\citenamefont{Prinz}(1998)}]{Pri98}
\bibinfo{author}{\bibfnamefont{G.~A.} \bibnamefont{Prinz}},
  \bibinfo{journal}{Science} \textbf{\bibinfo{volume}{282}},
  \bibinfo{pages}{1660} (\bibinfo{year}{1998}).

\bibitem[{\citenamefont{Wolf et~al.}(2001)\citenamefont{Wolf, Awschalom,
  Buhrman, Daughton, von Moln{\'a}r, Roukes, Chtchelkanova, and
  Treger}}]{Wol01}
\bibinfo{author}{\bibfnamefont{S.~A.} \bibnamefont{Wolf}},
  \bibinfo{author}{\bibfnamefont{D.~D.} \bibnamefont{Awschalom}},
  \bibinfo{author}{\bibfnamefont{R.~A.} \bibnamefont{Buhrman}},
  \bibinfo{author}{\bibfnamefont{J.~M.} \bibnamefont{Daughton}},
  \bibinfo{author}{\bibfnamefont{S.}~\bibnamefont{von Moln{\'a}r}},
  \bibinfo{author}{\bibfnamefont{M.~L.} \bibnamefont{Roukes}},
  \bibinfo{author}{\bibfnamefont{A.~Y.} \bibnamefont{Chtchelkanova}},
  \bibnamefont{and} \bibinfo{author}{\bibfnamefont{D.~M.}
  \bibnamefont{Treger}}, \bibinfo{journal}{Science}
  \textbf{\bibinfo{volume}{294}}, \bibinfo{pages}{1488} (\bibinfo{year}{2001}).

\bibitem[{\citenamefont{Datta and Das}(1990)}]{Dat90}
\bibinfo{author}{\bibfnamefont{D.}~\bibnamefont{Datta}} \bibnamefont{and}
  \bibinfo{author}{\bibfnamefont{B.}~\bibnamefont{Das}},
  \bibinfo{journal}{Appl.~Phys.~Lett.} \textbf{\bibinfo{volume}{56}},
  \bibinfo{pages}{665} (\bibinfo{year}{1990}).

\bibitem[{\citenamefont{Butler et~al.}(1997)\citenamefont{Butler, Zhang, Wang,
  van Ek, and MacLaren}}]{But97}
\bibinfo{author}{\bibfnamefont{W.~H.} \bibnamefont{Butler}},
  \bibinfo{author}{\bibfnamefont{X.-G.} \bibnamefont{Zhang}},
  \bibinfo{author}{\bibfnamefont{X.}~\bibnamefont{Wang}},
  \bibinfo{author}{\bibfnamefont{J.}~\bibnamefont{van Ek}}, \bibnamefont{and}
  \bibinfo{author}{\bibfnamefont{J.~M.} \bibnamefont{MacLaren}},
  \bibinfo{journal}{J.~Appl.~Phys.} \textbf{\bibinfo{volume}{81}},
  \bibinfo{pages}{5518} (\bibinfo{year}{1997}).

\bibitem[{\citenamefont{MacLaren et~al.}(1998)\citenamefont{MacLaren, Butler,
  and Zhang}}]{Mac98}
\bibinfo{author}{\bibfnamefont{J.~M.} \bibnamefont{MacLaren}},
  \bibinfo{author}{\bibfnamefont{W.~H.} \bibnamefont{Butler}},
  \bibnamefont{and} \bibinfo{author}{\bibfnamefont{X.-G.} \bibnamefont{Zhang}},
  \bibinfo{journal}{J.~Appl.~Phys.} \textbf{\bibinfo{volume}{83}},
  \bibinfo{pages}{6521} (\bibinfo{year}{1998}).

\bibitem[{\citenamefont{Ko\v{s}uth et~al.}(2000)\citenamefont{Ko\v{s}uth,
  Min{\'a}r, Cabria, Perlov, Crisan, Ebert, and Akai}}]{Kos00}
\bibinfo{author}{\bibfnamefont{M.}~\bibnamefont{Ko\v{s}uth}},
  \bibinfo{author}{\bibfnamefont{J.}~\bibnamefont{Min{\'a}r}},
  \bibinfo{author}{\bibfnamefont{I.}~\bibnamefont{Cabria}},
  \bibinfo{author}{\bibfnamefont{A.~Y.} \bibnamefont{Perlov}},
  \bibinfo{author}{\bibfnamefont{V.}~\bibnamefont{Crisan}},
  \bibinfo{author}{\bibfnamefont{H.}~\bibnamefont{Ebert}}, \bibnamefont{and}
  \bibinfo{author}{\bibfnamefont{H.}~\bibnamefont{Akai}}, in
  \emph{\bibinfo{booktitle}{Proceedings to the International symposium on
  structure and dynamics of heterogeneous systems, Duisburg, August 28--29}}
  (\bibinfo{year}{2000}).

\bibitem[{\citenamefont{Krebs et~al.}(1987)\citenamefont{Krebs, Jonker, and
  Prinz}}]{Kre87}
\bibinfo{author}{\bibfnamefont{J.~J.} \bibnamefont{Krebs}},
  \bibinfo{author}{\bibfnamefont{B.~T.} \bibnamefont{Jonker}},
  \bibnamefont{and} \bibinfo{author}{\bibfnamefont{G.~A.} \bibnamefont{Prinz}},
  \bibinfo{journal}{Appl.~Phys.~Lett.} \textbf{\bibinfo{volume}{61}},
  \bibinfo{pages}{2526} (\bibinfo{year}{1987}).

\bibitem[{\citenamefont{Reiger et~al.}(2000)\citenamefont{Reiger, Reinwald,
  Garreau, Ernst, Z{\"o}lfl, Bensch, Bauer, Preis, and Bayreuther}}]{Rei00}
\bibinfo{author}{\bibfnamefont{E.}~\bibnamefont{Reiger}},
  \bibinfo{author}{\bibfnamefont{E.}~\bibnamefont{Reinwald}},
  \bibinfo{author}{\bibfnamefont{G.}~\bibnamefont{Garreau}},
  \bibinfo{author}{\bibfnamefont{M.}~\bibnamefont{Ernst}},
  \bibinfo{author}{\bibfnamefont{M.}~\bibnamefont{Z{\"o}lfl}},
  \bibinfo{author}{\bibfnamefont{F.}~\bibnamefont{Bensch}},
  \bibinfo{author}{\bibfnamefont{S.}~\bibnamefont{Bauer}},
  \bibinfo{author}{\bibfnamefont{H.}~\bibnamefont{Preis}}, \bibnamefont{and}
  \bibinfo{author}{\bibfnamefont{G.}~\bibnamefont{Bayreuther}},
  \bibinfo{journal}{J.~Appl.~Phys.} \textbf{\bibinfo{volume}{87}},
  \bibinfo{pages}{5923} (\bibinfo{year}{2000}).

\bibitem[{\citenamefont{Bensch et~al.}(2001)\citenamefont{Bensch, Garreau,
  Moosb{\"u}hler, Beaurepaire, and Bayreuther}}]{Ben01}
\bibinfo{author}{\bibfnamefont{F.}~\bibnamefont{Bensch}},
  \bibinfo{author}{\bibfnamefont{G.}~\bibnamefont{Garreau}},
  \bibinfo{author}{\bibfnamefont{R.}~\bibnamefont{Moosb{\"u}hler}},
  \bibinfo{author}{\bibfnamefont{E.}~\bibnamefont{Beaurepaire}},
  \bibnamefont{and}
  \bibinfo{author}{\bibfnamefont{G.}~\bibnamefont{Bayreuther}},
  \bibinfo{journal}{J.~Appl.~Phys.} \textbf{\bibinfo{volume}{89}},
  \bibinfo{pages}{7133} (\bibinfo{year}{2001}).

\bibitem[{\citenamefont{Z{\"o}lfl et~al.}(1997)\citenamefont{Z{\"o}lfl,
  Brockmann, K{\"o}hler, Kreuzer, Schweinb{\"o}ck, Miethaner, Bensch, and
  Bayreuther}}]{Zol97}
\bibinfo{author}{\bibfnamefont{M.}~\bibnamefont{Z{\"o}lfl}},
  \bibinfo{author}{\bibfnamefont{M.}~\bibnamefont{Brockmann}},
  \bibinfo{author}{\bibfnamefont{M.}~\bibnamefont{K{\"o}hler}},
  \bibinfo{author}{\bibfnamefont{S.}~\bibnamefont{Kreuzer}},
  \bibinfo{author}{\bibfnamefont{T.}~\bibnamefont{Schweinb{\"o}ck}},
  \bibinfo{author}{\bibfnamefont{S.}~\bibnamefont{Miethaner}},
  \bibinfo{author}{\bibfnamefont{F.}~\bibnamefont{Bensch}}, \bibnamefont{and}
  \bibinfo{author}{\bibfnamefont{G.}~\bibnamefont{Bayreuther}},
  \bibinfo{journal}{J.~Magn.~Magn.~Mater.} \textbf{\bibinfo{volume}{175}},
  \bibinfo{pages}{16} (\bibinfo{year}{1997}).

\bibitem[{\citenamefont{MacLaren et~al.}(1990)\citenamefont{MacLaren, Crampin,
  Vvedensky, Albers, and Pendry}}]{Mac90}
\bibinfo{author}{\bibfnamefont{J.~M.} \bibnamefont{MacLaren}},
  \bibinfo{author}{\bibfnamefont{S.}~\bibnamefont{Crampin}},
  \bibinfo{author}{\bibfnamefont{D.~D.} \bibnamefont{Vvedensky}},
  \bibinfo{author}{\bibfnamefont{R.~C.} \bibnamefont{Albers}},
  \bibnamefont{and} \bibinfo{author}{\bibfnamefont{J.~B.}
  \bibnamefont{Pendry}}, \bibinfo{journal}{Comput.~Phys.~Commun.}
  \textbf{\bibinfo{volume}{60}}, \bibinfo{pages}{365} (\bibinfo{year}{1990}).

\bibitem[{\citenamefont{Ohtake et~al.}(1999)\citenamefont{Ohtake, Hanada,
  Yasuda, Arai, and Yao}}]{Oht99}
\bibinfo{author}{\bibfnamefont{A.}~\bibnamefont{Ohtake}},
  \bibinfo{author}{\bibfnamefont{T.}~\bibnamefont{Hanada}},
  \bibinfo{author}{\bibfnamefont{T.}~\bibnamefont{Yasuda}},
  \bibinfo{author}{\bibfnamefont{K.}~\bibnamefont{Arai}}, \bibnamefont{and}
  \bibinfo{author}{\bibfnamefont{T.}~\bibnamefont{Yao}},
  \bibinfo{journal}{Phys.~Rev.~B} \textbf{\bibinfo{volume}{60}},
  \bibinfo{pages}{8326} (\bibinfo{year}{1999}).

\bibitem[{\citenamefont{Mirbt et~al.}(1999)\citenamefont{Mirbt, Moll, Kley, and
  Joannopulos}}]{Mir99}
\bibinfo{author}{\bibfnamefont{S.}~\bibnamefont{Mirbt}},
  \bibinfo{author}{\bibfnamefont{N.}~\bibnamefont{Moll}},
  \bibinfo{author}{\bibfnamefont{A.}~\bibnamefont{Kley}}, \bibnamefont{and}
  \bibinfo{author}{\bibfnamefont{J.~D.} \bibnamefont{Joannopulos}},
  \bibinfo{journal}{Surf.~Sci.} \textbf{\bibinfo{volume}{422}},
  \bibinfo{pages}{L177} (\bibinfo{year}{1999}).

\bibitem[{\citenamefont{Harrison}(1979)}]{Har79}
\bibinfo{author}{\bibfnamefont{W.~A.} \bibnamefont{Harrison}},
  \bibinfo{journal}{J.~Vac.~Sci.~Technol.} \textbf{\bibinfo{volume}{16}},
  \bibinfo{pages}{1492} (\bibinfo{year}{1979}).

\bibitem[{\citenamefont{Hohenberg and Kohn}(1964)}]{Hoh64}
\bibinfo{author}{\bibfnamefont{P.}~\bibnamefont{Hohenberg}} \bibnamefont{and}
  \bibinfo{author}{\bibfnamefont{W.}~\bibnamefont{Kohn}},
  \bibinfo{journal}{Phys.~Rev.} \textbf{\bibinfo{volume}{136}},
  \bibinfo{pages}{B864} (\bibinfo{year}{1964}).

\bibitem[{\citenamefont{Kohn and Sham}(1965)}]{Koh65}
\bibinfo{author}{\bibfnamefont{W.}~\bibnamefont{Kohn}} \bibnamefont{and}
  \bibinfo{author}{\bibfnamefont{L.~J.} \bibnamefont{Sham}},
  \bibinfo{journal}{Phys.~Rev.} \textbf{\bibinfo{volume}{140}},
  \bibinfo{pages}{A1133} (\bibinfo{year}{1965}).

\bibitem[{\citenamefont{von Barth and Hedin}(1972)}]{Bar72}
\bibinfo{author}{\bibfnamefont{U.}~\bibnamefont{von Barth}} \bibnamefont{and}
  \bibinfo{author}{\bibfnamefont{L.}~\bibnamefont{Hedin}},
  \bibinfo{journal}{J.~Phys.~C} \textbf{\bibinfo{volume}{5}},
  \bibinfo{pages}{1629} (\bibinfo{year}{1972}).

\bibitem[{\citenamefont{Pant and Rajagopal}(1972)}]{Pan72}
\bibinfo{author}{\bibfnamefont{M.~M.} \bibnamefont{Pant}} \bibnamefont{and}
  \bibinfo{author}{\bibfnamefont{A.~K.} \bibnamefont{Rajagopal}},
  \bibinfo{journal}{Solid State Commun.} \textbf{\bibinfo{volume}{10}},
  \bibinfo{pages}{1157} (\bibinfo{year}{1972}).

\bibitem[{\citenamefont{Rajagopal and Callaway}(1973)}]{Raj73}
\bibinfo{author}{\bibfnamefont{A.~K.} \bibnamefont{Rajagopal}}
  \bibnamefont{and} \bibinfo{author}{\bibfnamefont{J.}~\bibnamefont{Callaway}},
  \bibinfo{journal}{Phys.~Rev.~B} \textbf{\bibinfo{volume}{7}},
  \bibinfo{pages}{1912} (\bibinfo{year}{1973}).

\bibitem[{\citenamefont{Hellwege and Hellwege}(1971)}]{LBIII6}
\bibinfo{editor}{\bibfnamefont{K.-H.} \bibnamefont{Hellwege}} \bibnamefont{and}
  \bibinfo{editor}{\bibfnamefont{A.~M.} \bibnamefont{Hellwege}}, eds.,
  \emph{\bibinfo{title}{Landolt-B{\"o}rnstein, Group III: Crystal and Solid
  State Physics}}, vol. \bibinfo{volume}{6: Structure Data of Elements and
  Intermetallic Phases} (\bibinfo{publisher}{Springer-Verlag, Berlin},
  \bibinfo{year}{1971}).

\bibitem[{\citenamefont{Hellwege and Hellwege}(1979)}]{LBIII11}
\bibinfo{editor}{\bibfnamefont{K.-H.} \bibnamefont{Hellwege}} \bibnamefont{and}
  \bibinfo{editor}{\bibfnamefont{A.~M.} \bibnamefont{Hellwege}}, eds.,
  \emph{\bibinfo{title}{Landolt-B{\"o}rnstein, Group III: Crystal and Solid
  State Physics}}, vol. \bibinfo{volume}{11: Elastic, Piezoelectric,
  Pyroelectric, Electrooptic Constants and Nonlinear Dielectric
  Susceptibilities of Crystals} (\bibinfo{publisher}{Springer-Verlag, Berlin},
  \bibinfo{year}{1979}).

\bibitem[{\citenamefont{Lee}(1973)}]{Lee73}
\bibinfo{author}{\bibfnamefont{B.~H.} \bibnamefont{Lee}},
  \bibinfo{journal}{J.~Appl.~Phys.} \textbf{\bibinfo{volume}{41}},
  \bibinfo{pages}{2988} (\bibinfo{year}{1973}).

\bibitem[{\citenamefont{Kresse and Hafner}(1993)}]{Kre93}
\bibinfo{author}{\bibfnamefont{G.}~\bibnamefont{Kresse}} \bibnamefont{and}
  \bibinfo{author}{\bibfnamefont{J.}~\bibnamefont{Hafner}},
  \bibinfo{journal}{Phys.~Rev.~B} \textbf{\bibinfo{volume}{47}},
  \bibinfo{pages}{558} (\bibinfo{year}{1993}).

\bibitem[{\citenamefont{Kresse and Furthm{\"u}ller}(1996)}]{Kre96}
\bibinfo{author}{\bibfnamefont{G.}~\bibnamefont{Kresse}} \bibnamefont{and}
  \bibinfo{author}{\bibfnamefont{J.}~\bibnamefont{Furthm{\"u}ller}},
  \bibinfo{journal}{Phys.~Rev.~B} \textbf{\bibinfo{volume}{54}},
  \bibinfo{pages}{11169} (\bibinfo{year}{1996}).

\bibitem[{VAS()}]{VASP}
\bibinfo{howpublished}{VASP homepage:
  \texttt{http://cms.mpi.univie.ac.at/vasp/}}.

\bibitem[{\citenamefont{Kresse and Hafner}(1994)}]{Kre94}
\bibinfo{author}{\bibfnamefont{G.}~\bibnamefont{Kresse}} \bibnamefont{and}
  \bibinfo{author}{\bibfnamefont{J.}~\bibnamefont{Hafner}},
  \bibinfo{journal}{J.~Phys.: Condens.~Matter} \textbf{\bibinfo{volume}{6}},
  \bibinfo{pages}{8245} (\bibinfo{year}{1994}).

\bibitem[{\citenamefont{Vanderbilt}(1990)}]{Van90}
\bibinfo{author}{\bibfnamefont{D.}~\bibnamefont{Vanderbilt}},
  \bibinfo{journal}{Phys.~Rev.~B} \textbf{\bibinfo{volume}{41}},
  \bibinfo{pages}{7892} (\bibinfo{year}{1990}).

\bibitem[{\citenamefont{Perdew and Zunger}(1981)}]{Per81}
\bibinfo{author}{\bibfnamefont{J.~P.} \bibnamefont{Perdew}} \bibnamefont{and}
  \bibinfo{author}{\bibfnamefont{A.}~\bibnamefont{Zunger}},
  \bibinfo{journal}{Phys.~Rev.~B} \textbf{\bibinfo{volume}{23}},
  \bibinfo{pages}{5048} (\bibinfo{year}{1981}).

\bibitem[{\citenamefont{Ceperley and Alder}(1980)}]{Cep80}
\bibinfo{author}{\bibfnamefont{D.~M.} \bibnamefont{Ceperley}} \bibnamefont{and}
  \bibinfo{author}{\bibfnamefont{B.~J.} \bibnamefont{Alder}},
  \bibinfo{journal}{Phys.~Rev.~Lett.} \textbf{\bibinfo{volume}{45}},
  \bibinfo{pages}{566} (\bibinfo{year}{1980}).

\bibitem[{\citenamefont{Perdew and Wang}(1992)}]{Per92}
\bibinfo{author}{\bibfnamefont{J.~P.} \bibnamefont{Perdew}} \bibnamefont{and}
  \bibinfo{author}{\bibfnamefont{Y.}~\bibnamefont{Wang}},
  \bibinfo{journal}{Phys.~Rev.~B} \textbf{\bibinfo{volume}{45}},
  \bibinfo{pages}{13244} (\bibinfo{year}{1992}).

\bibitem[{\citenamefont{Monkhorst and Pack}(1976)}]{Mon76}
\bibinfo{author}{\bibfnamefont{H.~J.} \bibnamefont{Monkhorst}}
  \bibnamefont{and} \bibinfo{author}{\bibfnamefont{J.~D.} \bibnamefont{Pack}},
  \bibinfo{journal}{Phys.~Rev.~B} \textbf{\bibinfo{volume}{13}},
  \bibinfo{pages}{5188} (\bibinfo{year}{1976}).

\bibitem[{\citenamefont{Murnaghan}(1944)}]{Mur44}
\bibinfo{author}{\bibfnamefont{F.~D.} \bibnamefont{Murnaghan}},
  \bibinfo{journal}{Proc.~Natl.~Acad.~Sci.} \textbf{\bibinfo{volume}{30}},
  \bibinfo{pages}{244} (\bibinfo{year}{1944}).

\bibitem[{\citenamefont{Zhu et~al.}(1992)\citenamefont{Zhu, Wang, and
  Louie}}]{Zhu92}
\bibinfo{author}{\bibfnamefont{J.}~\bibnamefont{Zhu}},
  \bibinfo{author}{\bibfnamefont{X.~W.} \bibnamefont{Wang}}, \bibnamefont{and}
  \bibinfo{author}{\bibfnamefont{S.~G.} \bibnamefont{Louie}},
  \bibinfo{journal}{Phys.~Rev.~B} \textbf{\bibinfo{volume}{45}},
  \bibinfo{pages}{8887} (\bibinfo{year}{1992}).

\bibitem[{\citenamefont{H{\"a}glund}(1993)}]{Hag93}
\bibinfo{author}{\bibfnamefont{J.}~\bibnamefont{H{\"a}glund}},
  \bibinfo{journal}{Phys.~Rev.~B} \textbf{\bibinfo{volume}{47}},
  \bibinfo{pages}{566} (\bibinfo{year}{1993}).

\bibitem[{\citenamefont{Lee et~al.}(1995)\citenamefont{Lee, Lee, and
  Ihm}}]{Lee95}
\bibinfo{author}{\bibfnamefont{G.}~\bibnamefont{Lee}},
  \bibinfo{author}{\bibfnamefont{M.~H.} \bibnamefont{Lee}}, \bibnamefont{and}
  \bibinfo{author}{\bibfnamefont{J.}~\bibnamefont{Ihm}},
  \bibinfo{journal}{Phys.~Rev. B} \textbf{\bibinfo{volume}{52}},
  \bibinfo{pages}{1459} (\bibinfo{year}{1995}).

\bibitem[{\citenamefont{Asato et~al.}(1999)\citenamefont{Asato, Settels,
  Hoshino, Asada, Zeller, and Dederichs}}]{Asa99}
\bibinfo{author}{\bibfnamefont{M.}~\bibnamefont{Asato}},
  \bibinfo{author}{\bibfnamefont{A.}~\bibnamefont{Settels}},
  \bibinfo{author}{\bibfnamefont{T.}~\bibnamefont{Hoshino}},
  \bibinfo{author}{\bibfnamefont{T.}~\bibnamefont{Asada}},
  \bibinfo{author}{\bibfnamefont{S.~B.~R.} \bibnamefont{Zeller}},
  \bibnamefont{and} \bibinfo{author}{\bibfnamefont{P.~H.}
  \bibnamefont{Dederichs}}, \bibinfo{journal}{Phys.~Rev.~B}
  \textbf{\bibinfo{volume}{60}}, \bibinfo{pages}{5202} (\bibinfo{year}{1999}).

\bibitem[{\citenamefont{Wijn}(1986)}]{LBIII19a}
\bibinfo{editor}{\bibfnamefont{H.~P.~J.} \bibnamefont{Wijn}}, ed.,
  \emph{\bibinfo{title}{Landolt-B{\"o}rnstein, Group III: Crystal and Solid
  State Physics}}, vol. \bibinfo{volume}{19: Magnetic Properties of Metals,
  Subvolume a: $3d$, $4d$, and $5d$ Elements, Alloys, and Compounds}
  (\bibinfo{publisher}{Springer-Verlag, Berlin}, \bibinfo{year}{1986}).

\bibitem[{\citenamefont{Bagayoko and Callaway}(1983)}]{Bag83}
\bibinfo{author}{\bibfnamefont{D.}~\bibnamefont{Bagayoko}} \bibnamefont{and}
  \bibinfo{author}{\bibfnamefont{J.}~\bibnamefont{Callaway}},
  \bibinfo{journal}{Phys.~Rev.~B} \textbf{\bibinfo{volume}{28}},
  \bibinfo{pages}{5419} (\bibinfo{year}{1983}).

\bibitem[{\citenamefont{Chelikowsky and Cohen}(1976)}]{Che76}
\bibinfo{author}{\bibfnamefont{J.~R.} \bibnamefont{Chelikowsky}}
  \bibnamefont{and} \bibinfo{author}{\bibfnamefont{M.~L.} \bibnamefont{Cohen}},
  \bibinfo{journal}{Phys.~Rev.~B} \textbf{\bibinfo{volume}{13}},
  \bibinfo{pages}{826} (\bibinfo{year}{1976}).

\bibitem[{\citenamefont{Payne et~al.}(1992)\citenamefont{Payne, Teter, Allan,
  Arias, and Joannopoulos}}]{Pay92}
\bibinfo{author}{\bibfnamefont{M.~C.} \bibnamefont{Payne}},
  \bibinfo{author}{\bibfnamefont{M.~P.} \bibnamefont{Teter}},
  \bibinfo{author}{\bibfnamefont{D.~C.} \bibnamefont{Allan}},
  \bibinfo{author}{\bibfnamefont{T.~A.} \bibnamefont{Arias}}, \bibnamefont{and}
  \bibinfo{author}{\bibfnamefont{J.~D.} \bibnamefont{Joannopoulos}},
  \bibinfo{journal}{Rev.~Mod.~Phys.} \textbf{\bibinfo{volume}{64}},
  \bibinfo{pages}{1045} (\bibinfo{year}{1992}).

\bibitem[{\citenamefont{Brailsford}(1966)}]{Bra66}
\bibinfo{author}{\bibfnamefont{F.}~\bibnamefont{Brailsford}},
  \emph{\bibinfo{title}{Physical Principles of Magnetism}}
  (\bibinfo{publisher}{D.~van Nostrand Company Ltd, London},
  \bibinfo{year}{1966}).

\bibitem[{\citenamefont{Madelung}(1978)}]{Mad78}
\bibinfo{author}{\bibfnamefont{O.}~\bibnamefont{Madelung}},
  \emph{\bibinfo{title}{Introduction to Solid-State Theory}}
  (\bibinfo{publisher}{Springer Verlag, Berlin}, \bibinfo{year}{1978}).

\end{thebibliography}

\end{document}